\documentstyle[aps,epsf,epsfig]{revtex}

\newcommand{\bq}{\begin{equation}}
\newcommand{\eq}{\end{equation}}
\newcommand{\bqn}{\begin{eqnarray}}
\newcommand{\eqn}{\end{eqnarray}}
\newcommand{\nb}{\nonumber}
\newcommand{\lb}{\label}

\begin{document}
\title{Kink Stability of  Self-Similar Solutions of Scalar Field in 
$2+1$ Gravity }
\author{  
Anzhong Wang  \thanks{E-mail: Anzhong$\_$Wang@baylor.edu}
${ }^{1}$, and Yumei Wu \thanks{ E-mail: yumei@im.ufrj.br. 
On leave from Institute of Mathematics, the Federal University of 
Rio de Janeiro,  Caixa Postal 68530,
   CEP 21945-970, Rio de Janeiro, RJ, Brazil} ${ }^{2}$}
\address{   
${ }^{1}$ CASPER, Department of Physics, Baylor University, Waco,
Texas 76798  \\
${ }^{2}$ Department of Mathematics, Baylor University, Waco, Texas 76798  }
\date{\today }

\maketitle

\begin{abstract}

The kink stability of  self-similar solutions of a massless scalar
field with circular symmetry in $2+1$ gravity  is studied, and
found that such solutions are unstable against the kink perturbations
along the sonic line (self-similar horizon). However, when
perturbations outside the sonic line are considered, and taking 
the ones along the sonic line as their boundary conditions, we find
that non-trivial perturbations do not exist.  In other words, the
consideration of perturbations outside the sonic line limits the
unstable mode of the perturbations found along the sonic line.   
As a result, the critical solution for the scalar collapse remains
critical even after the kink perturbations are taken into account.

\end{abstract}

\vspace{.7cm}


\vspace{.7cm}

\section{Introduction}

\renewcommand{\theequation}{1.\arabic{equation}}
\setcounter{equation}{0}

The studies of non-linearity of the Einstein field equations near
the threshold of black hole formation reveal very rich phenomena
\cite{Chop93}, which are quite similar to critical phenomena in
statistical mechanics and quantum field theory \cite{Golden}. In
particular, by numerically studying the gravitational collapse of
a massless scalar field in $3+1$-dimensional spherically symmetric
spacetimes, Choptuik found that the mass of such formed black
holes takes the form  \cite{Chop93},
 \bq
 \lb{1.1}
  M_{BH} = C(p)\left(p -p^{*}\right)^{\gamma},
 \eq
where $C(p)$ is a finite constant with $C(p^{*}) \not= 0 $, and
$p$ parameterizes a family of initial data in such a way that when
$p > p^{*}$  black holes are formed, and when $p < p^{*}$ no black
holes are formed. It was shown that, in contrast to $C(p)$, the
exponent $\gamma$   is {\em universal} to all the families of
initial data studied, and was numerically determined as $\gamma
\sim 0.37$. The solution with $p = p^{*}$, usually called the
critical solution, is found also {\em universal}. Choptuik's
studies were soon generalized to other matter fields \cite{Gun00}.
From all the work done so far, the collapse in general falls into
two different  types, depending on whether the black hole mass
takes the scaling form (\ref{1.1}) or not. When it takes the form,
the corresponding collapse is called Type $II$ collapse, and when
it does not it is called Type $I$ collapse. In the type $II$
collapse, {\em all} the critical solutions found so far have
either discrete self-similarity (DSS) or homothetic
self-similarity (HSS), depending on the matter fields. In the type
$I$ collapse, the critical solutions have neither DSS nor HSS. For
certain matter fields, these two types of collapse can co-exist. A
critical solution in both two types has {\em one and only
one unstable mode}. This now is considered as one of the main
criteria for a solution to be critical.

The studies of critical collapse  have been mainly numerical so
far, and analytical ones are still highly hindered by the
complexity of the problem, even after imposing some symmetries.
Lately, some progress has been achieved in the studies of critical
collapse of a  scalar field in an anti-de Sitter
background in $2+1$-dimensional spacetimes both numerically
\cite{PC00,HO01} and analytically \cite{Gar01,CF01,GG02,HW02}.
This serves as the first analytical model in critical collapse. In
particular, Garfinkle \cite{Gar01} first found a class of exact
solutions to Einstein-scalar field equations, denoted by $S[n]$,
and later Garfinkle and Gundlach (GG) studied their linear 
perturbations and
found that the solution with $n = 2$ has only one unstable mode
\cite{GG02}. By definition this is a critical solution, and the
corresponding exponent $\gamma$ in Eq.(\ref{1.1}) can be read off
from the expression \cite{Even}
 \bq
 \lb{1.2}
 \gamma = \frac{1}{|k_{1}|},
 \eq
from which it was found $\gamma= 4/3$, where
$k_{1}$ denotes the unstable mode. Although the exponent $\gamma$
is close to that found numerically by Pretorius and Choptuik
\cite{PC00}, $\gamma \sim 1.2 \pm 0.05$ (but not to the one of Husain
and Olivier, $\gamma \sim 0.81$), this solution is different from
the numerical critical solution \cite{Gar01}. Using different
boundary conditions,   Hirschmann, Wang and Wu (HWW) found that
the solution with $n = 4$ has only one unstable mode \cite{HW02}.
As first noted by Garfinkle \cite{Gar01}, this $n = 4$ solution
matches extremely well with the numerical critical solution found
by Pretorius and Choptuik \cite{PC00}. However, the corresponding
exponent $\gamma$ now is given by $\gamma = |k_{1}|^{-1} = 4$,
which is significantly different from the numerical ones. The
boundary conditions used by HWW are \cite{HW02}: (a) The
perturbations must be free of spacetime singularity on the
symmetry axis; (b) They are analytical across the self-similarity
horizon, as the background solutions do; (c) No matter  field come
out of the already formed trapped region  \cite{Chandra83}. 
GG  considered only Conditions (a) and (b)  \cite{GG02}.

In this paper we shall study another important issue of critical
collapse for a scalar field, the kink stability. The kink modes 
result from the existence of critical characteristic lines (they are 
also referred to as self-similarity horizons, and sonic lines), 
along which discontinuities of
(higher order) derivatives of some  physical quantities can be
developed and propagate. The instability is characterized by the
divergence of the discontinuity, and the blow-up may imply the
formation of shock waves \cite{CF48}. An example that
discontinuities of derivatives can propagate along a sonic
line is given by the linear perturbation, $\delta\varphi(\tau,z) =
\varphi_{1}(z)e^{k\tau}$, of the massless scalar field in 
$2+1$ gravity, which satisfies the following  equation \cite{HW02},
 \bq
 \lb{lp}
 z\left(1-z\right){\varphi_{1}}'' 
 + \frac{1}{2}\left[(1+2k) - z(3+2k)\right]
 {\varphi_{1}}' - \frac{1}{2} k {\varphi_{1}} = f(z),
 \eq
where a prime denotes the ordinary differentiation with respect to
the indicated argument, $f(z)$ is a smooth function of $z$, and $z
= 1$ is the location of the sonic line  (cf. Eq.(112) in 
\cite{HW02}). From the above equation we can see that
it is possible for $\varphi_{1}$ to has discontinuous derivatives
only across the line $z =1$. In fact, assume that
$\varphi_{1}$ is continuous across $z = 1$ (but not its
first-order derivative), we can   write it in the form
 \bq
 \lb{lpa}
 \varphi_{1}(z) = \varphi_{1}^{+}(z)H(z-1)
 + \varphi_{1}^{-}(z)\left[1 - H(z-1)\right],
 \eq
where $H(x)$ denotes the Heavside (step) function, defined as
 \bq
 \lb{lpb}
 H(x) = \cases{1, & $x > 0$,\cr
 0, & $x < 0$.\cr}
 \eq
Then, we find that
 \bqn
 \lb{lpc}
 {\varphi_{1}}' &=& {
 \varphi_{1}^{+}}' H(z-1)
 + {\varphi_{1}^{-}}'\left[1 - H(z-1)\right],\nb\\
{\varphi_{1}}'' &=& {\varphi_{1}^{+}}'' H(z-1)
 + {\varphi_{1}^{-}}''\left[1 - H(z-1)\right] +
 \left[{\varphi_{1}}'\right]^{-}\delta(z-1),
 \eqn
where $\delta(x)$ denotes the Dirac delta function, and
 \bq
 \lb{lpd}
 \left[{\varphi_{1}}'\right]^{-} \equiv
 \lim_{z \rightarrow 1^{+0}}{\left(\frac{d\varphi_{1}^{+}(z)}{dz}\right)}
  - \lim_{z \rightarrow 1^{-0}}{\left(\frac{d\varphi_{1}^{-}(z)}{dz}\right)}.
  \eq
Substituting Eq.(\ref{lpc}) into Eq.(\ref{lp}) and considering the
facts
 \bqn
 \lb{lpe}
 & & H^{m}(x) = H(x),\;\;\; \left[1 - H(x)\right]^{m} = \left[1 -
 H(x)\right],\nb\\
& &  \left[1 - H(x)\right] H(x) =0,\;\; x \delta(x) = 0,
 \eqn
where $m$ is an integer, one can see that Eq.(\ref{lp}) holds also
on the horizon $z =1$ even when $\left[{\varphi_{1}}'\right]^{-}
\not= 0$. This is because   $(1-z)\left[{\varphi_{1}}'\right]^{-}
\delta(z-1) = 0$, as long as $\left[{\varphi_{1}}'\right]^{-}$ is finite.
 Thus, we have
\bq
\lb{lpea}
\left[{\delta\varphi}_{,z}\right]^{-} = 
\left[{\varphi_{1}}'\right]^{-} e^{k\tau},
\eq
where $(\;)_{,z} \equiv \partial (\;)/\partial z$.
The above expression shows clearly how the discontinuity of the first 
derivative of the perturbation
$\delta\varphi(\tau,z)$ propagate along the sonic line $z = 1$. When
$Re(k) > 0$ the perturbation grows exponentially as $\tau \rightarrow 
\infty$, and is said unstable with respect to the kink perturbation. 
When $Re(k) < 0$ the perturbation decays exponentially and is said stable.

Note that if the discontinuity happened on other places,
say, $z = z_{0} \not= 1$, clearly Eq.(\ref{lp}) would not hold on
$z = z_{0}$, because now $(1-z)\left[{\varphi_{1}}'\right]^{-}
\delta(z-z_{0})\not= 0$. This explains why the discontinuities are
allowed only along the sonic lines.
The above analysis also shows that the kink perturbations are
different from the ones considered in \cite{GG02} and \cite{HW02},
because there it was required that $\varphi_{1}(z)$ is analytical
across $z = 1$, that is,
 \bq
 \lb{lpf}
 \left[{\varphi_{1}}^{(m)}\right]^{-} = 0,\;\; (m = 1, 2, ...)
 \eq
where ${\varphi_{1}}^{(m)}$ denotes the $m$-th order derivatives of
$\varphi_{1}$. Therefore, kink perturbations were excluded in the
studies of linear perturbations of \cite{GG02} and \cite{HW02}.

Ori and Piran first studied kink stability of self-similar
solutions in newtonian gravity \cite{OP88}, and lately Harada
generalized such a study to the relativistic case and found that
the critical self-similar solutions of a perfect fluid with the
equation of state $P = k \rho$  are not stable against kink
perturbations for $k \ge 0.89$, where $P$ and $\rho$ denote,
respectively, the pressure and energy density of the fluid
\cite{Har01}.  More recently, Harada and Maeda  showed that in
four-dimensional spherically symmetric case the self-similar
massless scalar solution found lately by Brady {\em et al}
\cite{Brady02} is also not stable against kink perturbations
\cite{HM03}. 

In this paper, we study the kink stability of the scalar field
in $2+1$ gravity. Instead of assuming that $\delta\varphi(\tau,z)$
is $c^{0}$ across the sonic line, as we did in the above example, 
following Harada \cite{Har01}, and   Harada and Maeda \cite{HM03},
we shall assume that $\delta\varphi(\tau,z)$ is $c^{1}$,
that is, $\delta\varphi(\tau,z)$ and its first-order derivative with
respect to $z$ are continuous across the sonic line, but not its 
second-order derivative. We shall first show that perturbations
obtained along the sonic line allow the existence of unstable 
modes. However, when we consider perturbations outside the sonic 
line and take the ones obtained along the sonic line as their boundary
conditions, we find that these conditions together with the ones on the
symmetry axis do not allow any non-trivial perturbations in the regions
outside the sonic line. Therefore,
the consideration of perturbations in the whole spacetime limits
the unstable mode found along the sonic line. Thus,  
{\em all the self-similar solutions of the
massless scalar field are stable against kink perturbations in
$2+1$ gravity}. As a result, the critical solution for the scalar
collapse remains critical, even after the kink perturbations are
taken into account. 

Specifically,   the paper is organized
as follows: In Sec. II we give a brief review of the self-similar
solution, which is needed in the studies of linear perturbations
in Sec. III, in which we first consider the linear perturbations of
the self-similar solutions along the sonic line, and then the perturbations
outside the sonic line. In Sec. IV, we summarize the main results obtained
in this paper and then present our concluding remarks.

\section{The Einstein-Scalar Field Equations  }

\renewcommand{\theequation}{2.\arabic{equation}}
\setcounter{equation}{0}

The general form of metric for a ($2+1$)-dimensional spacetime with
circular  symmetry can be cast in the form,
 \bq
 \lb{2.1}
 ds^2 = - 2 e^{2\sigma(u,v)} du dv + r^2(u,v) d\theta^2,
 \eq
where $(u,v)$ is a pair of null coordinates varying in the range
$(-\infty,\infty)$, and $\theta$ is the usual angular coordinate 
with the hypersurfaces $\theta = 0, \; 2\pi$ being identified. 
$\xi_{(\theta)} = \partial_{\theta}$ is a Killing vector. 
It should be noted that the form of the metric is unchanged under the 
coordinate transformations,
 \bq
 \lb{2.2}
 u = u(\bar{u}),\;\;\;\; v = v(\bar{v}).
 \eq
To have circular symmetry, some  conditions on the symmetry axis
needed to be imposed. In 
general this is not trivial. As a matter of fact,  only when the 
axis is free of spacetime singularity, do we know how to 
impose these conditions. Since in this paper we are mainly interested in 
gravitational collapse, we shall assume that the axis is regular  
at the beginning of the collapse. In particular,  we impose the following 
conditions:

(i) There must exist a symmetry axis, which can be expressed as  
\bq
\lb{2.3}
X \equiv \left|\xi^{\mu}_{(\theta)}\xi^{\nu}_{(\theta)}g_{\mu\nu} 
\right|  \rightarrow 0,
\eq
as $v \rightarrow f(u)$, where we assumed that the axis  is located at 
$r(v =f(u), u) = 0$.

(ii) The spacetime near the symmetry axis is locally flat, which can be
written as 
\bq
\lb{2.4}
\frac{X_{,\alpha}X_{,\beta} g^{\alpha\beta}}{4X} \rightarrow
 1,
\eq
as $v \rightarrow f(u)$, where $(\;)_{,\alpha} \equiv \partial (\;)/\partial 
x^{\alpha}$.

The corresponding Einstein-scalar field equations  for the metric (\ref{2.1}) 
take the form,
\bqn
\lb{2.7a}
 r_{,uu} - 2\sigma_{,u} r_{,u} &=& - 8\pi G r \phi_{,u}^{2},\\
\lb{2.7b}
 r_{,vv} - 2\sigma_{,v} r_{,v} &=& - 8\pi G r \phi_{,v}^{2},\\
\lb{2.7c}
 r_{,uv} + 2r\sigma_{,uv} &=& - 8\pi G r  \phi_{,u}\phi_{,v},\\
 \lb{2.7}
 r_{,uv} &=& 0,
 \eqn
while the equation of motion for the scalar field  is given by
 \bq
 \lb{2.8}
  2\phi_{,uv} + \frac{1}{r}\left(r_{,u}\phi_{,v} + r_{,v}\phi_{,u}\right)
 = 0.
 \eq  
To study self-similar solutions, we first introduce the dimensionless 
variables, $z$ and $\tau$, via the relations
 \bq
 \lb{2.9}
 z = \frac{v}{(-u)},\;\;\;\;
 \tau = - \ln\left(\frac{(-u)}{u_{0}}\right),
 \eq
where $u_{0}$ is a dimensional constant with the dimension of length, 
and the above relations are
assumed to be valid only in the region $v \ge 0,\; u \le 0$. We will refer 
to this region as Region $I$ [cf. Fig. 1].  

Self-similar solutions are given 
by
\bq
\lb{2.12}
F(\tau, z) = F_{ss}(z),
\eq
where $F \equiv \{\sigma, \; s, \; \varphi\}$, and
 \bqn
 \lb{2.11}
 r(u,v) &\equiv& (- u) s(\tau, z),\nb\\
 \phi(u,v) &\equiv& c \ln\left|-u\right| + \varphi(\tau, z),
 \eqn
with $c$ being an arbitrary constant. 
A class of such solutions was first found by Garfinkle \cite{Gar01}, 
which can be written as  \cite{HW02}
 \bqn
 \lb{2.15}
  \sigma_{ss}(u,v) &=& \frac{1}{2}
  \ln\left\{\frac{\left[v^{1/2} + \epsilon
 (-u)^{1/2}\right]^{4\chi}}{(- u v)^{\chi}}\right\}
 +  \sigma^{1}_{0}, \nb\\
 r_{ss}(u,v) &=&   (-u) - v, \nb\\
 \phi_{ss}(u,v)&=&  
 2c\ln\left|v^{1/2} + \epsilon (-u)^{1/2}\right| + \phi^{1}_{0},
 \eqn
where $\epsilon = \pm 1$, $\sigma^{1}_{0}$ and $\varphi^{1}_{0}$ 
are integration constants, and $\chi \equiv 8\pi G \, c^{2}$.
 As shown in \cite{HW02}, the hypersurface $v = 0$ for the solutions with
$1 > \chi \ge 1/2$ represents a sonic line, and the solutions can be 
extended across the hypersurface, whereby they  can be interpreted as 
representing the gravitational collapse of a scalar field, in which a black 
hole is finally formed. The extension can be realized by  introducing  
two new coordinates $\bar{u}$ and $\bar{v}$ via the relations
 \bq
 \lb{2.16}
 \bar{u} = - (-{u})^{1/2n},\;\;\;\;\;
 \bar{v}= v^{1/2n},
 \eq
 where $ n \equiv {1}/[2(1 - \chi)] \ge 1$.
In order to have the extension unique, we require that it be analytical 
across the hypersurface $v = 0$, which, in turn, requires $n$  to 
be an integer and satisfy the condition,
 \bq
 \lb{2.18}
 n = \frac{1}{2(1 - \chi)} = \cases{2l, & $\epsilon = 1$,\cr
 2l +1, & $\epsilon = - 1$,\cr}
 \eq
where $l$ is another integer. For the detail, we refer readers to
\cite{HW02}. In these new coordinates, the metric and the massless 
scalar field are given by
 \bqn
 \lb{2.19}
 ds^{2} & = & -  2e^{2\bar{\sigma}_{ss}(\bar{u},\bar{v})} \, d\bar{u}d\bar{v}
  + {r_{ss}}^{2}(\bar{u},\bar{v}) d\theta^{2}, \nb\\
  \bar{\sigma}_{ss}(\bar{u},\bar{v}) &=& \sigma_{ss}(u,v) 
  + \frac{1}{2}\ln\left\{4n^{2}
  \left(-\bar{u}\bar{v}\right)^{2n-1}\right\}\nb\\
   &=& \frac{1}{2}\ln\left\{4n^{2}
  \left|f(\bar{u},\bar{v})\right|^{4\chi}\right\} + \sigma^{1}_{0},\nb\\
  r_{ss}(\bar{u},\bar{v}) & = &  (- \bar{u})^{2n}
  - \bar{v}^{2n} \nb\\
  \phi_{ss}(\bar{u},\bar{v}) & = & 2c\ln\left| f(\bar{u},\bar{v}) \right|
  + \phi^{1}_{0},
 \eqn
where $ f(\bar{u},\bar{v}) \equiv  \bar{v}^{n} + \epsilon (-\bar{u})^{n}$.
Note that the symmetry axis (the vertical line $r = 0$ in Fig. 1)
is located at $\bar{v} = \bar{u}$, for which conditions (\ref{2.3}) requires
$\sigma^{1}_{0}  = \frac{1}{2}(1-4\chi)\ln(2)$.
The corresponding Penrose diagram is given by Fig. 1.

\begin{figure}[htbp]
 \begin{center}
 \label{fig1}
 \leavevmode
  \epsfig{file=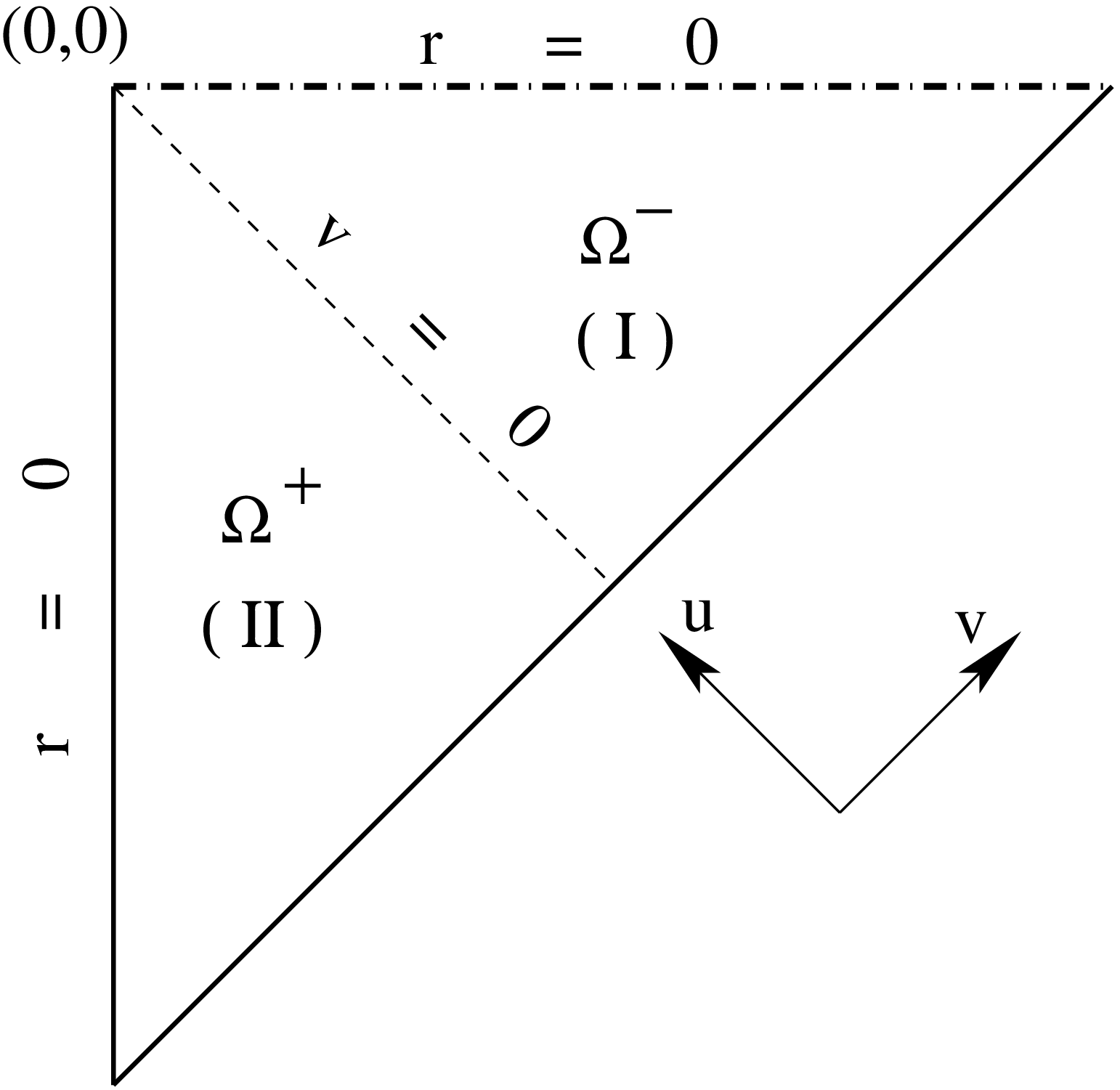,width=0.45\textwidth,angle=0}
 \caption{The Penrose diagram for the solutions given by 
 Eq.(\ref{2.19}).  $\phi_{,\alpha}$ is timelike in   Region $II$ and  null on 
the hypersurface $v = 0$. In Region I it is spacelike for $n = 2l$ and
timelike for $n = 2l +1$. The horizontal line $r = 0$ is singular for
$n = 2l +1$ but not for $n = 2l$. In Region $I$ all the rings 
of $u, \; v = Constant$ are trapped, while in Region $II$ none of them is
trapped.  }
 \end{center}
 \end{figure}

\section{  Linear Perturbations of Self-Similar Solutions: Kink Stability}
\renewcommand{\theequation}{3.\arabic{equation}}
\setcounter{equation}{0}

In this section, we consider the linear perturbations of the self-similar
solutions given by Eq.(\ref{2.19}). For the sake of
simplicity, we shall drop all the bars from $\bar{\sigma},\; \bar{u}$ 
and $\bar{v}$, so that the background solutions can be written as,
 \bqn
 \lb{3.1}
 ds^{2} &=& -2e^{2{\sigma}_{ss}(u,v)} du dv 
 + r_{ss}^{2}(u,v) d\theta^{2},\nb\\
{\sigma}_{ss}(u,v) &=& \frac{1}{2}\ln\left(4n^{2}
 \left|f(u, v)\right|^{4\chi}\right) + \sigma^{1}_{0},\nb\\
r_{ss}(u,v) & = &  (-u)^{2n} -  v^{2n}, \nb\\
\phi_{ss}(u,v) & = & 2c\ln\left| f(u, v)\right| + \phi^{1}_{0},\nb\\
f({u}, {v}) &\equiv&   {v}^{n} + \epsilon (-u)^{n}.
\eqn
Let us first divide the spacetime in Fig. 1 into three different
regions, $\Omega^{\pm}$ and $\Sigma$, defined, respectively, by
$\Omega^{+} = \left\{x^{\alpha}: u,\; v \le 0, \; v \ge u\right\}$, 
$\Omega^{-} = \left\{x^{\alpha}: u \le 0,\; v \ge 0,\; v \le |u|\right\}$, 
and $ \Sigma = \left\{x^{\alpha}:  v = 0\right\}$.
Then, for any given $C^{1}$ function $f(u,v)$, we can write it as
\bq
\lb{3.3}
f(u,v) = f^{+}(u,v)\left[1 - H(v)\right] + f^{-}(u,v)H(v),
\eq
where  $f^{\pm}$ denote the functions, defined, respectively, in the regions
$\Omega^{\pm}$. In the present case, we have
\bq
\lb{3.5}
f^{\pm}(u,v) = f^{\pm}_{ss}(u,v) + \delta{f^{\pm}}(u,v),
\eq
where $f_{ss}\equiv \{{\sigma}_{ss}, \; r_{ss},\; 
\phi_{ss}\}$ denotes the background solutions given by Eq.(\ref{3.1}),
which are analytical across $v = 0$,
\bq
\lb{3.6}
\lim_{v \rightarrow 0^{-}}
{\frac{\partial^{m} f^{+}_{ss}(u, v)}{\partial{v}^{m}}}
= \lim_{v \rightarrow 0^{+}}
{\frac{\partial^{m} f^{-}_{ss}(u, v)}{\partial{v}^{m}}},
\;\; (m = 0, 1, 2, ...).
\eq
Since $f(u,v)$ is $C^{1}$, we must have
\bqn
\lb{3.7}
\lim_{v \rightarrow 0^{-}}{\delta{f^{+}(u, v)}}
= \lim_{v \rightarrow 0^{+}}{\delta{f^{-}(u, v)}}
\equiv \delta{f_{c}(u)},\nb\\
\lim_{v \rightarrow 0^{-}}{\delta{f^{+}_{,v}(u, v)}}
= \lim_{v \rightarrow 0^{+}}{\delta{f^{-}_{,v}(u, v)}}
\equiv \delta{f^{(1)}_{c}(u)}.
\eqn
Then, we find
\bqn
\lb{3.8}
f_{,v}(u,v) &=& f^{+}_{,v}(u,v)\left[1 - H(v)\right] 
                  + f^{-}_{,v}(u,v)H(v),\nb\\
f_{,vv}(u,v) &=& f^{+}_{,vv}(u,v)\left[1 - H(v)\right] 
                  + f^{-}_{,vv}(u,v)H(v).
\eqn
Inserting Eqs.(\ref{3.5})-(\ref{3.8}) into Eqs.(\ref{2.7a})-(\ref{2.8}) 
and considering Eq.(\ref{lpe}), to the first order of $\delta{f}$, we obtain
\bqn
\lb{3.11a}
& & \delta{r}_{,uu} - 2\left(\sigma_{ss,u}\delta{r}_{,u}
+ r_{ss,u}\delta{\sigma}_{,u}\right)
= -8\pi G\left(2r_{ss}\phi_{ss,u}\delta\phi_{,u}
+  {\phi_{ss,u}^{2}}\delta{r}\right),\\
\lb{3.11b}
& & \delta{r}_{,vv} - 2\left(\sigma_{ss,v}\delta{r}_{,v}
+ r_{ss,v}\delta{\sigma}_{,v}\right)
= -8\pi G\left(2r_{ss}\phi_{ss,v}\delta\phi_{,v}
+ {\phi_{ss,v}^{2}}\delta{r}\right),\\
\lb{3.11c}
& &   2\left(r_{ss}\delta{\sigma}_{,uv} + \sigma_{ss,uv}\delta{r}\right)
= -8\pi G\left[r_{ss}\left(\phi_{ss,u}\delta\phi_{,v} 
    + \phi_{ss,v}\delta\phi_{,u}\right) 
 +  \phi_{ss,u} \phi_{ss,v} \delta{r}\right],\\
\lb{3.11d}
& & \delta{r}_{,uv} = 0,\\
\lb{3.11e}
& &  2r_{ss}\delta{\phi}_{,uv} + 2 {\phi}_{ss,uv}\delta{r} 
   + r_{ss,u}\delta{\phi}_{,v} + r_{ss,v}\delta{\phi}_{,u}
   + \phi_{ss,u}\delta{r}_{,v} + \phi_{ss,v}\delta{r}_{,u} = 0,
\eqn
where the quantities $f_{ss}$ and $\delta{f}$ should be understood as
$f^{+}_{ss}$ and $\delta{f^{+}}$ in $\Omega^{+}$, and $f^{-}_{ss}$ 
and $\delta{f^{-}}$ in $\Omega^{-}$.

\subsection{Kink Stability}

Kink stability is the study of the linear perturbations of 
Eqs.(\ref{3.11a}) - (\ref{3.11e}) along the sonic line $v = 0$.
To solve these equations for $\delta f_{c}(u)$, following
Ori and Piran \cite{OP88} (See also \cite{Har01,HM03}), 
we  impose the following conditions: 
Assume that the perturbations turn on at the moment, say,
$u = u_{0}$, then we require

(A)  the perturbations initially vanish in the interior,
\bq
\lb{3.12a}
\delta f^{-}(u_{0}, v) = 0,\; v \in \Omega^{-},
\eq

 (B)  the perturbations and their first-order derivatives
 be continuous everywhere, and in particular across the sonic line,  
\bq
\lb{3.12b}
\left[\delta f\right]^{-} = 0, \;\;\; 
\left[\delta f_{,v}\right]^{-} = 0, \left(v = 0\right),  
\eq

(C)  ${\delta \phi^{\pm}}_{,vv}$ and ${\delta \sigma^{\pm}}_{,vv}$  
be discontinuous across the sonic line, 
\bq 
\lb{3.12c}
\delta\phi''_{c}(u) \equiv \left[\delta \phi_{,vv} \right]^{-} \not= 0, \;\;\;
\delta\sigma''_{c}(u) \equiv \left[\delta \sigma_{,vv} \right]^{-} \not= 0, 
\;\; \left(v = 0\right).
\eq
From the above we first note that Eq.(\ref{3.12a}) remains true for all
$u > u_{0}$. In fact,  $\delta f^{-}(u, v) = 0$ are indeed solutions of   
Eqs.(\ref{3.11a}) - (\ref{3.11e}) in $\Omega^{-}$. Then, 
from Eqs.(\ref{3.12a}) and (\ref{3.12b}) we find 
\bqn
\lb{3.13}
& & \delta{f^{+}}(u, 0) = 0, \;\;\; \delta{f^{+}_{,v}}(u, 0) = 0,\nb\\
& & \delta\phi''_{c}(u) = \delta{\phi^{+}_{,vv}}(u, 0),\;\;\;
\delta\sigma''_{c}(u) = \delta{\sigma^{+}_{,vv}}(u, 0).
\eqn
Taking the limit $v \rightarrow 0^{-}$ in Eqs.(\ref{3.11a})-(\ref{3.11e}) and 
considering the above equation we find that
\bq
\lb{3.14}
\left[\delta{r}_{,vv}\right]^{-} = \delta{r^{+}_{,vv}}(u, 0) =  0.
\eq
On the other hand, taking  derivatives of Eqs.(\ref{3.11e})
and (\ref{3.11c}) with respect to $v$, and then taking the limit
$v \rightarrow 0^{-}$, we obtain
\bqn
\lb{3.15a}
& & 2r_{ss}\left(\delta\phi''_{c}\right)_{,u} + r_{ss,u}\delta\phi''_{c}
= 0, \\
\lb{3.15b}
& &  \left(\delta\sigma''_{c}\right)_{,u}  
= -4\pi G \phi_{ss,u} \delta\phi''_{c}, 
\eqn
along the sonic line $v = 0$. Substituting Eq.(\ref{3.1}) into the above 
equations and then integrating them, we obtain
\bqn
\lb{3.16}
\delta\phi''_{c}(u) &=& \frac{A}{(-u)^{n}} = \frac{A}{u^{1/2}_{0}}e^{\tau/2},\nb\\
 \delta\sigma''_{c}(u) &=& - \frac{8\pi G c A}{(-u)^{n}}
 = \frac{8\pi G c A}{u^{1/2}_{0}}e^{\tau/2},
\eqn
where $A$ is an integration constant. Since $n \ge 1$, from
the above expressions we can see that both $\delta\phi''_{c}(u)$ and 
$\delta\sigma''_{c}(u)$  diverge as $u \rightarrow 0^{-}\; 
({\mbox{or}} \; \tau \rightarrow \infty)$, or in other words,
the self-similar solutions are not stable against the kink perturbations.

It should be noted that $\delta{f}^{+}(u,v)$ cannot be zero identically in  
$\Omega^{+}$,  because we already have $\delta{f}^{-}(u,v) = 0$ in $\Omega^{-}$
and 
\bq 
\lb{3.16a}
\delta{f}''_{c}(u) = \delta{f}^{+}_{,vv}\left(u, 0^{-}\right) \not= 0.
\eq
Then, a natural question rises: Do the perturbations given by 
Eq.(\ref{3.16}) match to the ones in region $\Omega^{+}$? To answer this
question, in the next subsection we shall consider the linear perturbations
of Eqs.(\ref{3.11a})-(\ref{3.11e}) in region $\Omega^{+}$, by considering
Eqs.(\ref{3.13}), (\ref{3.14}) and (\ref{3.16}) as their boundary
conditions at $v = 0$.

\subsection{Linear Perturbations in $\Omega^{+}$}

To study the linear perturbations in $\Omega^{+}$, it
is found convenient to use the dimensionless variables $\tau$ and $z$,
defined by Eq.(\ref{2.9}). However, they are valid only in region 
$\Omega^{-}$. In region $\Omega^{+}$ we define them as
\bq
\lb{3.17}
\tilde{\tau} = - \ln\left(\frac{-\tilde{u}}{u_{0}}\right), \;\;\;
\tilde{z} = \frac{\tilde{v}}{\tilde{u}},
\eq
where $\tilde{u}, \; \tilde{v} \le 0$ in $\Omega^{+}$, and 
\bq
\lb{3.18}
\tilde{u} \equiv - (-{u})^{2n}, \;\;\;  \tilde{v} \equiv - (-{v})^{2n}.
\eq
The null coordinates $u$ and $v$ in Eq.(\ref{3.18}) should be
understood as the ones, $\bar{u}$ and $\bar{v}$, defined by Eq.(\ref{2.16}).
In terms of $\tilde{\tau}$ and $\tilde{z}$, the background solutions 
  (\ref{3.1}) in $\Omega^{+}$ can be written in the form,
\bqn
\lb{3.19}
s_{0}(\tilde{z}) &=& 1 - \tilde{z},\nb\\
\sigma_{0}(\tilde{z}) &=& 2\chi\ln\left(\tilde{z}^{1/4} + 
          \tilde{z}^{-1/4}\right) + \sigma^{1}_{0},\nb\\
\varphi_{0}(\tilde{z}) &=& 2c\ln\left(1 + \tilde{z}^{1/2}\right) 
    + \varphi^{1}_{0},
\eqn
with
\bqn
\lb{3.20}
\sigma_{ss}({u}, {v}) &=& \sigma_{0}(\tilde{z}) 
+ \frac{1}{2}\ln\left[4n^{2}
(-{u}{v})^{2n-1}\right],\nb\\
r_{ss}({u}, {v}) &=& (-{u})^{2n}s_{0}(\tilde{z}),\nb\\
\phi_{ss}({u}, {v}) &=& 
\varphi_{0}(\tilde{z}) + c\ln\left[(-{u})^{2n}\right].
\eqn
Again, $u$ and $v$ in Eq.(\ref{3.20}) should be
understood as $\bar{u}$ and $\bar{v}$ defined by Eq.(\ref{2.16}).
For detail, we refer readers to Eqs.(57)-(59) in \cite{HW02}. Without
causing any confusions, in the following we shall drop the tildes from 
$\tilde{\tau}$ and $\tilde{z}$. Then, writing the perturbations as
\bqn
\lb{3.21}
\delta{r} &=& (-\tilde{u})s_{1}({z})e^{k\tau},\nb\\
\delta{\sigma} &=& \sigma_{1}({z})e^{k\tau},\nb\\
\delta{\phi} &=& \varphi_{1}({z})e^{k\tau},
\eqn
it can be shown that the linearized perturbations given by  
Eqs.(\ref{3.11a})-(\ref{3.11e}) reduce exactly to the ones of
(67)-(71) of \cite{HW02}, the general solutions of which are
 Eqs.(110)-(118) for $k = 1$, and  Eqs.(120)-(125) for
$k \not= 1$, given in \cite{HW02}. In particular, $s_{1}(z)$ is given by
\bq
\lb{3.21a}
s_{1}(z) = \cases{\beta \ln(z) + s^{0}_{1}, & $ k = 1$,\cr
                   \beta z^{1-k} + s^{0}_{1}, & $ k \not= 1$,\cr}
\eq
where $\beta$ and $s^{0}_{1}$ are the integration constants.

However, since here we consider
the kink stability, the boundary conditions are different from the ones
used in \cite{GG02,HW02}. In particular, in \cite{GG02,HW02} it was 
required that the perturbations be analytical across the surface $v = 0$, 
while in the present case these conditions should be replaced by 
Eqs.(\ref{3.13}), (\ref{3.14}) and (\ref{3.16}), which can be written as
\bqn
\lb{3.22a}
k &=& \frac{1}{2},\nb\\
s_{1}(z) &\simeq&   O\left(z^{3}\right),\nb\\
\sigma_{1}(z) &\simeq& - \frac{4\pi{G}c A}{2u^{1/2}_{0}} z^{2} 
+ O\left(z^{3}\right),\nb\\
\varphi_{1}(z) &\simeq& \frac{A}{2u^{1/2}_{0}} z^{2} + O\left(z^{3}\right), 
\eqn
as $z \rightarrow 0$. Thus, the instability of the perturbations along the
hypersurface $v = 0$ found in the last subsection
is due to a single mode, $k = 1/2$. In addition to the 
above conditions, we also need to impose  some conditions on the symmetry
axis $r = 0$, so that the local-flatness conditions (\ref{2.3}) and 
(\ref{2.4}) are satisfied. In terms of $f_{1}(z)$, these conditions are 
exactly the ones given by Eq.(105) in \cite{HW02},
\bqn
 \lb{3.23}
 \left. s_{1}(z)\right|_{z=1} & = & 0, \nb\\
 \left. \sigma_{1}(z) \right|_{z=1} & \sim & \; {\rm finite} \nb\\
 \left. \left\{(1 - z)\frac{d\varphi_{1}(z)}{d z} - 2k 
 \varphi_{1}(z) \right\} \,
 \right|_{z=1} & \sim & \; {\rm finite},\;\; (r = 0).
 \eqn
  
From Eqs.(\ref{3.21a})  and (\ref{3.23}) we find that 
$s^{0}_{1} = - \beta$, while  
Eq.(\ref{3.22a})   requires $\beta = 0$,
for which the solutions of $\sigma_{1}(z)$ and $\varphi_{1}(z)$ 
with $k = 1/2$ are given by \cite{HW02}
\bqn
 \lb{3.25a}
 \sigma_{1}(z) &=& \frac{2\chi}{c}(1 - z^{1/2})\left[z^{1/2}(1 + z^{1/2})
  \frac{d\varphi_{1}(z)}{dz} + \frac{1}{2} \varphi_{1}\right],\\
 \lb{3.25b}
\varphi_{1} (z) &=& c_{1} \, F\left(\frac{1}{2}, \frac{1}{2}; 1; z\right)
  + c_{2} \, F\left(\frac{1}{2}, \frac{1}{2}; 1; 1-z\right),
 \eqn
where $c_{1}$ and $c_{2}$ are two arbitrary constants, 
and $F\left(a, b; c; z\right)$ 
denotes the ordinary hypergeometric function with 
$F\left(a, b; c; 0\right) = 1$. 
From  the expression \cite{AS72},
\bq
\lb{3.26}
F\left(\frac{1}{2}, \frac{1}{2}; 1; z\right) = \frac{1}{\pi^{2}}
\sum_{n=0}^{\infty}{\frac{2\Gamma^{2}\left(\frac{1}{2} + n\right)}
{\left(n!\right)^{2}} \left\{\psi(n+1) - \psi\left(n+\frac{1}{2}\right) 
- \frac{1}{2}\ln(1-z)\right\}(1-z)^{n}},
\eq
we find  
\bq
\lb{3.27}
F\left(\frac{1}{2}, \frac{1}{2}; 1; z\right) \rightarrow 
- \frac{1}{\pi}\ln(1-z),
\eq
as $z \rightarrow 1$. Then, Eq.(\ref{3.23}) requires $
c_{1} = 0$.
On the other hand, from Eq.(\ref{3.26}) we also find  
\bq
\lb{3.29}
F\left(\frac{1}{2}, \frac{1}{2}; 1; 1-z\right) \rightarrow 
- \frac{1}{\pi}\ln(z),
\eq
as $z \rightarrow 0$. Thus, the conditions of Eq.(\ref{3.22a}) yield 
$c_{2} = 0$. In review of all the above, 
we find that the boundary conditions 
(\ref{3.22a}) and (\ref{3.23}) require
\bq
\lb{3.30}
s_{1}(z) = \sigma_{1}(z) = \varphi_{1}(z) = 0.
\eq
That is,   non-trivial perturbations in $\Omega^{+}$ are not allowed by
the boundary conditions (\ref{3.22a}) and (\ref{3.23}). Then, we must
have $\delta{f}_{c}''(u) = 0$. In other words, 
{\em the consideration of the perturbations  in $\Omega^{+}$ limits 
the unstable mode of the perturbations along the sonic line $v = 0$}.

\section{Conclusions}

In this paper, we have studied the kink stability 
of the self-similar solutions
of a massless scalar field in $2+1$ gravity, 
and found that perturbations along
the sonic line (self-similar horizon) indeed 
allow the existence of an unstable 
mode. 

In the study of kink stability, it is assumed 
that the spacetime
inside the sonic line is not perturbed, that is, $\delta{f}^{-}(u,v) = 0$ 
identically \cite{OP88,Har01,HM03}. Then,   $\delta{f}^{+}(u,v)$ must be
non-zero outside the sonic line, in order to have non-vanishing perturbations 
along the sonic line. However, the perturbations outside the sonic line cannot 
be arbitrary. In particular, they have to match to the ones along the sonic 
line. In addition, they need also  satisfy some physical/geometrical conditions,
such as, the local-flatness conditions on the symmetry axis. A natural
question now is: After considering all these, does the spectrum of the 
perturbations obtained along the sonic line still remain the same?

To answer this question, in Sec. III we have studied the perturbations outside 
the sonic line, by taking the ones obtained along the sonic line as their
boundary conditions. We have shown explicitly that these 
conditions, together with the ones on the symmetry axis, indeed alter the
spectrum of the perturbations along the sonic line, and in particular, they
limit all the unstable modes. Thus, {\em all
the self-similar solutions of the massless scalar field in $2+1$ gravity is
stable against kink perturbations}. As a result, {\em the critical solution for the 
scalar collapse remains critical even after the kink perturbations are taken 
into account}.

Finally, we note that in the newtonian gravity the spectrum
of the perturbations along the sonic line remains the same, even after
the perturbations outside the sonic line are taken into account \cite{WW05}. 
It would be very interesting to see if this is still the case in 
four-dimensional spacetimes in the framework of Einstein's theory 
of gravity.

 
\section*{Acknowledgments}

One of the authors (AW) would like to express his gratitude to  M.W. Choptuik, 
D. Garfinkle, and T. Harada  for their valuable discussions and suggestions. 
Part of the work was done when one of the authors (YW) was visiting the 
Department of Mathematics, Baylor University. She would like to express 
her gratitude to the Department for hospitality. The authors also thank 
the Astrophysics Center, Zhejiang University of Technology for hospitality.

\end{document}